\documentclass[11pt,a4paper]{article}

\usepackage{jheppub}

\usepackage{graphicx} % Required for inserting images

\usepackage{amsmath}
\usepackage{amscd}
\usepackage{amsfonts}
\usepackage{amssymb}
\usepackage{mathrsfs}
\usepackage{fancybox} 
\usepackage{enumerate}
\usepackage[boxsize=0.8ex]{ytableau} %young diagram
\usepackage{youngtab}
\usepackage{graphicx}

\usepackage{color}
\usepackage{comment}

\def\={\stackrel{\bullet}{=}}

\def\({\left(}
\def\){\right)}
\def\[{\left[}
\def\]{\right]}

\def \be {\begin{equation}}
\def \ee {\end{equation}}
\def \beqa {\begin{eqnarray}}
\def \eeqa {\end{eqnarray}}
\def \beal#1 {\begin{align}#1\end{align}}
\def \bes#1 {\begin{equation}\begin{split}#1\end{split}\end{equation}}

\def\ket#1{|#1 \rangle}

\def\x't{(\boldsymbol{x'},t)}

\def\3tensor#1#2#3#4{#1^{#2\;#4}_{\;\;#3}}
\def\i{\mathrm{i}}

\def\KS{\text{KS}}
\def\OV{\text{OV}}
\def\W{\text{W}}
\begin{document}

%\begin{titlepage}
\vspace{-2cm}
\begin{flushright}
\normalsize{ 
%YITP-23-66
 OU-HET-1271}
\end{flushright}

\title{  Quantized Axial Charge 
in the Hamiltonian Approach to Wilson Fermions} 
\affiliation{Department of Physics, The University of Osaka, Toyonaka, Osaka 560-0043, Japan}
\author{ 
Tatsuya Yamaoka\footnote{\it t\_yamaoka@het.phys.sci.osaka-u.ac.jp}\,}

\abstract{
    We investigate the Hamiltonian formulation of 1+1~D
    staggered fermions and reconstruct vector and axial charge operators, found by Arkya Chatterjee et al.,
    using the Wilson fermion formalism.
    These operators commute with the Hamiltonian and become the generators of vector 
    and axial $\mathrm{U}(1)$ symmetries in the continuum limit. 
    An interesting feature of the axial charge operator is
    that it acts locally on operators and has quantized eigenvalues in momentum space. 
    Therefore, the eigenstates of this operator can be interpreted as fermion states with a well-defined integer chirality,
    analogous to those in the continuum theory. 
    This allows us to realize a gauge theory where the axial $\mathrm{U}(1)_A$ symmetry acts as a gauge symmetry.
    We construct a Hamiltonian using the eigenstates of the axial charge operator, 
    preserving exact axial symmetry on the lattice and vector symmetry in the continuum.
    As applications, we examine the implementation of the Symmetric Mass Generation (SMG) mechanism
    in the $1^4(-1)^4$ and 3-4-5-0 models.
    Our formulation supports symmetry-preserving interactions with quantized chiral charges,
    although further numerical studies are required to verify the SMG mechanism in interacting models.
}

\date{today}

\maketitle

%\newpage
\section{Introduction}
\label{intro}

The formulation of chiral gauge theories on the lattice remains one of the longstanding unsolved problems in the quantum field theory.
The fundamental obstruction stems from the Nielsen--Ninomiya's no-go theorem~\cite{Nielsen:1980rz,Nielsen:1981hk}, 
which states that any local, Hermitian, and translationally invariant lattice fermion action inevitably produces fermion doublers 
due to the periodicity of the Brillouin zone. 
Eliminating these doublers necessarily breaks chiral symmetry.

In the path-integral formalism, significant progress has been made in circumventing this problem. 
One of the most promising approaches is the overlap fermion formalism~\cite{Neuberger:1998wv}, 
based on the Ginsparg--Wilson relation~\cite{Ginsparg:1981bj}. 
Overlap fermions preserve a modified version of chiral symmetry at finite lattice spacing, 
enabling the description of certain chiral phenomena on the lattice~\cite{JLQCD:2008jiv,Aoki:2012yj,Cossu:2013uua,Hasenfratz:1998ri,Fujikawa:1998if,Suzuki:1998yz,Luscher:1998pqa,Adams:1998eg}.

In contrast, the corresponding Hamiltonian formulation of chiral symmetry on the lattice is still under active development. 
Analogous to the Ginsparg--Wilson relation, a Dirac operator has been proposed in the Hamiltonian formalism, 
which allow for the construction of axial charge operator that commute with 
both the Hamiltonian and a vector charge operator~\cite{Horvath:1998gq,Hayata:2023skf,Hayata:2023zuk}. 
However, this axial charge operator is momentum-dependent and lack quantized eigenvalues, 
making it difficult to define well-behaved chiral projections over the entire momentum space.

Recently, Ref.~\cite{Chatterjee:2024gje} proposed a new formulation for 1+1~D staggered fermions, 
where both the vector and axial charge operators are local, commute with the Hamiltonian, 
and possess quantized eigenvalues\footnote{The quantized axial charge was first found by Refs.~\cite{Thacker:1994ns,Horvath:1998gq} in terms of the connection to integrability.}. These charges generate distinct $\mathrm{U}(1)$ symmetries 
that can be gauged individually. 
Based on this structure, one can construct well-defined chiral fermions.
Moreover, the noncommutativity between the vector and axial charges encodes the chiral anomaly. 
Further developments include Ref.~\cite{Xu:2025hfs},
where a mechanism for the cancellation of chiral anomalies involving $Q_V$ and $Q_A$ was proposed.

In this paper, we focus on the observation that the Hamiltonian of staggered fermions in 1+1~D can be smoothly deformed into that of Wilson fermions. 
We reinterpret the structure of the axial charge operator proposed in Ref.~\cite{Chatterjee:2024gje} 
using the Wilson fermion formulation. The eigenstates of the defined axial charge are constructed 
as linear combinations of positive-energy creation and negative-energy annihilation operators.
The resulting Hamiltonian includes particle-number non-conserving terms. 
Interestingly, we find that this Hamiltonian still admits conserved charges associated 
with vector $\mathrm{U}(1)$ symmetry in the continuum limit, 
which remain noncommutative with the axial charge and therefore do not conflict with the Nielsen--Ninomiya's theorem.

The resulting 1+1~D Hamiltonian, constructed via this formulation, 
proves useful for building chiral gauge theories using the symmetric mass generation (SMG) mechanism~\cite{Wang:2013yta,Wang:2018ugf}. 
SMG refers to the mechanism by which interactions, without introducing fermion bilinear terms, 
gap out a system while preserving symmetries.
Importantly, SMG requires the underlying theory to be anomaly-free. 
In the presence of 't~Hooft anomalies, 
the interaction terms must explicitly break the anomalous symmetry.
Ref.~\cite{Xu:2025hfs} explored the 3-4-5-0 chiral model~\cite{Wang:2013yta,Wang:2018ugf,Wang:2022fzc,Zeng:2022grc,Onogi:2025tev} 
and proposed that its anomaly-free symmetries can be constructed using $Q_V$ and $Q_A$, with anomaly cancellation occurring between the two. 
However, their proposed interaction term fails to commute with one of the symmetry generators except at special momenta, 
thus violating the assumed symmetry. 
This suggests that the two symmetries are trivially anomaly-free rather 
than mutually anomaly-canceling on the lattice. In this paper, 
we aim to reconstruct the 3-4-5-0 model based on this insight.

This paper is organized as follows. 
In Sec.~\ref{sec: Equivalence-Stg-OV-fermions}, 
we demonstrate the equivalence between the staggered and Wilson fermion Hamiltonians in 1+1~D. 
In Sec.~\ref{sec:Axial-charge-Staggered-fermions}, 
we briefly review the two lattice-quantized charge operators that commute with the staggered fermion Hamiltonian, 
following Ref.~\cite{Chatterjee:2024gje}. 
In Sec.~\ref{sec:Eigenstates-Axial-charge}, 
we redefine the axial charge operator $Q_A$ using Wilson fermions and construct the Hamiltonian in terms of the fields with definite charge of $Q_A$. 
We verify that this Hamiltonian still supports two conserved charges corresponding to vector and axial $\mathrm{U}(1)$ symmetries in the continuum limit.
In Sec.~\ref{sec:SMG}, 
we examine the applicability of the SMG mechanism in the presence of gauged $\mathrm{U}(1)_A$ symmetry generated by $Q_A$, 
using the $1^4(-1)^4$ model~\cite{Kikukawa:2017gvk} and the 3-4-5-0 model as examples.
We summarize our paper in Sec.~\ref{sec:Conclusion}.
In Appendix~\ref{sec:Onsager}, 
we briefly review the structure of the Onsager algebra. 
In Appendix~\ref{OV-Wilson-Chiralcharge}, we show that the 1+1~D Hamiltonian of Wilson fermions can be deformed into that of overlap fermions. 
In particular, we demonstrate that the Wilson-Dirac operator satisfies the Ginsparg--Wilson relation, 
and that the chiral operator defined through this Dirac operator corresponds to the unquantized axial charge built from Onsager algebra charges.

%%%%%%%%%%%%%%%%%%%%%%%%%%%%%%%%%%%%%%%%%%%%%%%%%%%%
%%%%%%%%%%%%%%%%%%%%%%%%%%%%%%%%%%%%%%%%%%%%%%%%%%%%
\section{Equivalence of Staggered and Wilson Fermions in $1+1$ Dimensions}
\label{sec: Equivalence-Stg-OV-fermions}
In this section, we see that the Hamiltonian of staggered fermions and that of Wilson fermions are equivalent in 1+1~D, massless, and non-interacting case~\cite{Hayata:2023skf}.
Let us start with a Hamiltonian of the 1+1 dimensional staggered 
fermion~\cite{Kogut:1974ag,Banks:1975gq,Susskind:1976jm,Catterall:2025vrx},  
\begin{align}
    \label{eq:Def:Staggered-Hamiltonian}
    H_{\KS} &= \i \sum_{j=1}^{2N} \(c_j^\dagger c_{j+1} + c_j c_{j+1}^\dagger\) \notag \\  
    &= \frac{\i}{2} \sum_{j=1}^{2N} \(a_{j} a_{j+1} + b_{j}b_{j+1}\) \ ,
\end{align}
where $c_j$ denotes a single-component complex fermion satisfying the anticommutation relation  
\begin{align}
    \{c_j , c_{j'}^\dagger\} = \delta_{j,j'} \ .
\end{align}
The complex fermion $c_j$ can be decomposed into two Majorana fermions $a_j$ and $b_j$, satisfying  
\begin{align}
    a_j = a_j^\dagger, \quad b_j = b_j^\dagger, \quad \{a_{j}, a_{j'}\} = \{b_{j}, b_{j'}\} = 2 \delta_{jj'} \ ,
\end{align}
as follows,
\begin{align}
    c_j = \frac{1}{2} \( a_j + b_j \) \ .
\end{align}
In the continuum limit, 
the Hamiltonian~\eqref{eq:Def:Staggered-Hamiltonian} 
describes a single free massless Dirac fermion.  

By introducing a two-component fermion $\psi_j = \( \psi_{L,j}, \psi_{R,j} \)^T = \( c_{2j}, \i c_{2j+1} \)^T$,  
and defining the gamma matrices as  
\begin{align}
    \label{eq:Def:E-GammaMatrices-sigma-1}
    \gamma^0 = \sigma^1 \ , \quad \gamma^1 = -\sigma^3 \ , \quad \gamma^5 = - \i \gamma^0 \gamma^1 = \sigma^2 \ ,
\end{align}
the Hamiltonian can be rewritten as  
\begin{align}
    H_\KS &= \sum_{j=1}^{N} \psi^\dagger_j \gamma^0 \[ \gamma^1 \frac{1}{2}\(\nabla + \nabla^\dagger \) - \frac{1}{2} \nabla^\dagger \nabla  \] \psi_j \notag \\  
    &=: \sum_{j=1}^{N} \psi^\dagger_j \gamma^0 \mathcal{D}_W(0)  \psi_j := H_\W \ ,
\end{align}
where the discrete derivatives are defined as  
\begin{align}
    \label{Def:differentiation}
    \nabla \psi_j &= \psi_{j+1} - \psi_{j} \ , \quad  
    \nabla^\dagger \psi_j = \psi_{j} - \psi_{j-1} \ , \notag \\  
    \nabla^\dagger \nabla \psi_j &= \psi_{j+1} + \psi_{j-1} - 2 \psi_j \ .
\end{align}
Here, $\mathcal{D}_W(m)$ is the Wilson--Dirac operator, given by  
\begin{equation}
    \label{Wilson-E}
    \mathcal{D}_W (m)= \frac{1}{2} \{ \gamma^1 (\nabla + \nabla^\dagger) - \nabla^\dagger \nabla \} - m \ .
\end{equation}
Thus, in 1+1~D, 
the Hamiltonian of the staggered fermion $H_\KS$ is equivalent to 
the Hamiltonian of the Wilson fermion $H_\W$.

%%%%%%%%%%%%%%%%%%%%%%%%%%%%%%%%%%%%%%%%%%%%%%%%%%%%
%%%%%%%%%%%%%%%%%%%%%%%%%%%%%%%%%%%%%%%%%%%%%%%%%%%%
\section{Conserved Charges in the Hamiltonian of Staggered Fermions}
\label{sec:Axial-charge-Staggered-fermions}
In a 1+1~D lattice Hamiltonian of staggered fermions with a finite-dimensional Hilbert space,
it has been pointed out that there exist two conserved charge operators~\cite{Chatterjee:2024gje,Thacker:1994ns,Horvath:1998gq},
 $Q_V$ and $Q_A$, each of which corresponds to the vector or axial $\mathrm{U}(1)$ symmetry in the continuum limit.  
Both $Q_V$ and $Q_A$ act locally on operators and can be gauged independently, which is why they serve as generators of the $\mathrm{U}(1)$ symmetry.  
Interestingly, on a finite lattice, these charges do not commute and instead generate the Onsager algebra, while their commutator vanishes in the continuum limit.  
This non-Abelian algebra accounts for the chiral anomaly, in accordance with the Nielsen-Ninomiya theorem.  
Here, we review the properties of the charge operators $Q_V$ and $Q_A$.

The conserved charge operators $Q_V$ and $Q_A$ are defined as
\begin{align}
    \label{eq:Def:vector-operator}
    Q_V &= \sum_{j=1}^{2N} \(c_j ^\dagger c_j - \frac{1}{2} \)   = \frac{\i}{2} \sum_{j=1}^{2N} a_j b_j  \equiv \sum_{j=1}^{2N}q_j ^V \ , \\
    \label{eq:Def:axial-operator}
    Q_A &= \frac{1}{2} \sum_{j=1}^{2N} \( c_j + c_j ^\dagger \) \(c_{j+1} - c_{j+1}^\dagger \)   = \frac{\i}{2} \sum_{j=1}^{2N} a_j b_{j+1}  \equiv \sum_{j=1}^{2N} q_{j+\frac{1}{2}} ^A \ ,
\end{align}
both of which are quantized and commute with the Hamiltonian $H_\KS$,
\begin{align}
    \[H_\KS, Q_V\]=\[H_\KS, Q_A\]=0 \ .
\end{align}
However, interestingly, they do not commute themselves,
\begin{align}
    \label{eq:noncommutativity-QVQA}
    \[Q_V,Q_A\] &= - \sum_{j=1}^{2N} \(c_j c_{j+1} + c_{j}^\dagger c_{j+1}^\dagger \)   \\
    &  =i G_1 \notag \\
    & \neq 0  \ , \notag
\end{align}
which vanishes in the continuum limit. 
The term $G_1$ in the second line in Eq.\eqref{eq:noncommutativity-QVQA} is a generator of the Onsager algebra whose definition are described in Appendix~\ref{sec:Onsager}.
This noncommutative property implies the existence of a mixed anomaly between vector and axial $\mathrm{U}(1)$ symmetries.

%%%%%%%%%%%%%%%%%%%%%%%%%%%%%%%%%%%%%%%%%%%%%%%%%%%%
%%%%%%%%%%%%%%%%%%%%%%%%%%%%%%%%%%%%%%%%%%%%%%%%%%%%

\section{Eigenstates of the Axial charge and the Hamiltonian}
\label{sec:Eigenstates-Axial-charge}
We have seen
that the staggered fermion system with staggered Hamiltonian can be rewritten 
in terms of the Wilson fermion system with Wilson Hamiltonian.
Consequently, the conserved vector and axial charge operators $Q_V$ and $Q_A$, originally defined 
for the free massless staggered fermion Hamiltonian in Eqs.~\eqref{eq:Def:vector-operator} and \eqref{eq:Def:axial-operator}, 
can naturally be redefined as conserved charges commuting with the Wilson fermion Hamiltonian. 
An interesting feature of the axial charge operator $Q_A$ is that it is an on-site symmetry and has quantized eigenvalues in momentum space. Therefore, the eigenstates of this operator can be interpreted as fermion states with a well-defined integer chirality, analogous to those in the continuum theory. This allows us to construct a gauge theory where the axial $\mathrm{U}(1)_A$ symmetry acts as a gauge symmetry.

To achieve the above purpose, we first construct the Hamiltonian in coordinate space based on the eigenstates of the axial charge operator given in Eq.~\eqref{eq:Def:axial-operator}. 
The fermions defined as follows are eigenstates of the axial charge operator with integer eigenvalues $\pm 1$,
\begin{align}
    \label{eq:Def:Axial-fermion-Coordinate-Space-Pi/4-dagger}
    \begin{pmatrix}
        \Psi_{L,j}^{\dagger} \\
        \Psi_{R,j}^{\dagger} 
    \end{pmatrix}
    &= \frac{1}{2\sqrt{2}} 
    \begin{pmatrix}
        -2c_{2j}^\dagger + (c_{2j+1} - c_{2j+1}^\dagger) - (c_{2j-1} + c_{2j-1}^\dagger) \\
         2c_{2j}^\dagger + (c_{2j+1} - c_{2j+1}^\dagger) - (c_{2j-1} + c_{2j-1}^\dagger)
    \end{pmatrix} \ .
\end{align}
These fermions satisfy the commutation relations
\begin{align}
    [Q_A, \Psi_{L,j}^{\dagger}] = \Psi_{L,j}^{\dagger}, \quad 
    [Q_A, \Psi_{R,j}^{\dagger}] = -\Psi_{R,j}^{\dagger}.
\end{align}
It is straightforward to verify that these fermions satisfy the canonical quantization conditions.
Using these fermions, the Hamiltonian $H_\W$ can be rewritten as
\begin{align}
    \label{eq:Hamiltonian-PsiH-position}
    H_{\W} = \i \sum_{j=1}^{N} \biggl[ 
        &\Psi_{L,j}^{\dagger} \frac{1}{2}(\nabla + \nabla^\dagger) \Psi_{L,j}^{} 
        - \Psi_{R,j}^{\dagger} \frac{1}{2}(\nabla + \nabla^\dagger) \Psi_{R,j}^{} \notag  \\
        & - \Psi_{R,j}^{} \frac{1}{2} \nabla\nabla^\dagger \Psi_{L,j}^{} 
        - \Psi_{L,j}^{\dagger} \frac{1}{2} \nabla\nabla^\dagger \Psi_{R,j}^{\dagger}
    \biggr] \ ,
\end{align}
where $\nabla$ and $\nabla^\dagger$ are the forward and backward difference operators defined in Eq.~\eqref{Def:differentiation}. 
The second term in Eq.~\eqref{eq:Hamiltonian-PsiH-position} corresponds to the Wilson mass term, which explicitly breaks the vector $\mathrm{U}(1)$ symmetry associated with fermion number conservation. However, it is important to emphasize that this Hamiltonian still possesses conserved charge operators, $Q_V$ and $Q_A$, which flow to the vector and axial $\mathrm{U}(1)$ charges in the continuum limit. 
In this fermionic basis, these charges are given by
\begin{align}
    Q_V &= \frac{1}{2} \sum_{j=1}^{N} \biggl[ 
        \Psi_{L,j}^{\dagger} \Psi_{L,j}^{} + \Psi_{R,j}^{\dagger} \Psi_{R,j}^{} 
        - \Psi_{L,j}^{\dagger} \Psi_{R,j}^{} - \Psi_{R,j}^{\dagger} \Psi_{L,j}^{} -2 \notag \\
    & - \frac{1}{2} 
        \biggl\{ (\Psi_{L,j}^{} -  \Psi_{L,j}^{\dagger} + \Psi_{R,j}^{} -  \Psi_{R,j}^{\dagger})  
        (\Psi_{L,j+1}^{} +  \Psi_{L,j+1}^{\dagger} + \Psi_{R,j+1}^{} +  \Psi_{R,j+1}^{\dagger})  
        \biggr\}
    \biggr] \ , 
\end{align}
\begin{align}
    Q_A &= \sum_{j=1}^{N} (\Psi_{L,j}^{\dagger} \Psi_{L,j}^{} - \Psi_{R,j}^{\dagger} \Psi_{R,j}^{}) \ .
\end{align}
By construction, these satisfy $[H_W, Q_V] = [H_W, Q_A] = 0$. Furthermore, it follows that the eigenvalues of $Q_A$ on the fermion states $\Psi^{H}_{\alpha}$ are $\pm 1$.
Finally, the commutation relations of $Q_V$ with the fermionic operators are given by
\begin{align}
    [Q_V, \Psi_{\alpha,j}^{\dagger}] &= \Psi_{\alpha,j}^{\dagger} + \frac{1}{4} (\nabla + \nabla^\dagger)(\Psi_{L,j}^{} + \Psi_{R,j}^{}) + \nabla\nabla^\dagger(\Psi_{L,j}^{\dagger} + \Psi_{R,j}^{\dagger}), \\
    [Q_V, \Psi_{\alpha,j}^{}] &= -\Psi_{\alpha,j}^{} - \frac{1}{4} (\nabla + \nabla^\dagger)(\Psi_{L,j}^{\dagger} + \Psi_{R,j}^{\dagger}) - \nabla\nabla^\dagger(\Psi_{L,j}^{} + \Psi_{R,j}^{}).
\end{align}

Working in momentum space provides a clearer understanding of these concepts.  
The Fourier expansion of the one-component staggered fermion is given by  
\begin{align}
    c_j = \frac{1}{\sqrt{2N}} \sum \gamma_k e^{\frac{2\pi i}{2N}kj} \ .
\end{align}  
Thus, the Fourier modes of the fermions $\Psi_{\alpha}^H$ are expressed as  
\begin{align}
     \begin{pmatrix}
        \tilde{\Psi}_{L,j}^{\dagger} \\  
         \tilde{\Psi}_{R,j}^{\dagger}
     \end{pmatrix} &=  \frac{1}{2} \begin{pmatrix}
        \i  \sin \(\frac{2\pi}{2N}k\) \(- \gamma_{-k} + \gamma_{-k+N}\) + \gamma_{k}^{\dagger} \( -1 - \cos \(\frac{2\pi}{2N}k\)\) + \gamma_{k+N}^{\dagger} \( -1 + \cos\(\frac{2\pi}{2N}k\)\) \\  
       \i  \sin \(\frac{2\pi}{2N}k\) \(- \gamma_{-k} + \gamma_{-k+N}\) + \gamma_{k}^{\dagger} \(1 - \cos \(\frac{2\pi}{2N}k\)\) + \gamma_{k+N}^{\dagger} \(1 + \cos\(\frac{2\pi}{2N}k\)\)
   \end{pmatrix} \ .
\end{align}
Here, the first Brillouin zone is given by $ - \frac{N}{2} \leq k <  \frac{N}{2}$.  
The Hamiltonian can be expanded as follows:  
\begin{align}
     H_\W 
    %&= \sum_{- \frac{N}{2} \leq k < \frac{N}{2}} \biggl[  \(\tilde{\Psi}_{L,k}^{\dagger}\tilde{\Psi}_{L,k}^{} - \tilde{\Psi}_{R,k}^{\dagger}\tilde{\Psi}_{R,k}^{} \) \cos \(\frac{2\pi}{2N}k \) \sin \(\frac{2\pi}{2N}k \)  \notag \\  
    %  &\ \ \ \ \ \ \ \ \ \ \ \ \quad \quad \quad - \i \( \tilde{\Psi}_{L,k}^{\dagger} \tilde{\Psi}_{R,-k}^{\dagger} - \tilde{\Psi}_{R,-k}^{} \tilde{\Psi}_{L,k}^{}  \) \sin ^2  \(\frac{2\pi}{2N}k \)  \biggr] \notag \\  
     &=\frac{1}{2} \sum_{- \frac{N}{2} \leq k < \frac{N}{2}} \biggl[  \(\tilde{\Psi}_{L,k}^{\dagger}\tilde{\Psi}_{L,k}^{} - \tilde{\Psi}_{R,k}^{\dagger}\tilde{\Psi}_{R,k}^{} \)  \sin \(\frac{2\pi}{N}k \)  \notag \\  
     &\ \ \ \ \ \ \ \ \ \ \ \ \quad \quad \quad - \i \( \tilde{\Psi}_{L,k}^{\dagger} \tilde{\Psi}_{R,-k}^{\dagger} - \tilde{\Psi}_{R,-k}^{} \tilde{\Psi}_{L,k}^{}  \) \( 1-\cos  \(\frac{2\pi}{N}k \) \) \biggr] \ .
\end{align}
Moreover, the commutators between $Q_V$ and the fermion operators in momentum space are given by  
\begin{align}
     \[Q_V, \tilde{\Psi} _{\alpha,k}^{\dagger}\] &= \tilde{\Psi} _{\alpha,k}^{\dagger} + \frac{1}{2} \biggl\{ \i \sin \(\frac{2\pi}{N}k \) \( \tilde{\Psi} _{L,k}^{} +   \tilde{\Psi} _{R,k}^{}  \)  - \(1-\cos \(\frac{2\pi}{N}k \) \) \(\tilde{\Psi} _{L,k}^{\dagger} + \tilde{\Psi} _{R,k}^{\dagger}\)    \biggr\} \ , \\  
     \[Q_V, \tilde{\Psi} _{\alpha,k}^{}\] & = - \tilde{\Psi} _{\alpha,k}^{} + \frac{1}{2} \biggl\{ \i \sin \(\frac{2\pi}{N}k \) \( \tilde{\Psi} _{L,k}^{\dagger} +   \tilde{\Psi} _{R,k}^{\dagger}  \)  + \(1-\cos \(\frac{2\pi}{N}k \) \) \(\tilde{\Psi} _{L,k}^{} + \tilde{\Psi} _{R,k}^{}\)    \biggr\} \ .
\end{align}
Taking the limit $N \to \infty$ at finite momentum, we obtain  
\begin{align}
     \[Q_V, \tilde{\Psi}_{\alpha,k}^{\dagger}\] &=  \tilde{\Psi}_{\alpha,k}^{\dagger} \ , \\  
     \[Q_V, \tilde{\Psi} _{\alpha,k}^{}\] & = - \tilde{\Psi} _{\alpha,k}^{} \ , 
\end{align}
which confirms that, as expected, the charge operator $Q_V$ flows to the charge operator of the vector $\mathrm{U}(1)_V$ symmetry in the continuum limit.

From the above considerations, 
we have constructed a Hamiltonian using Weyl fermions with integer-valued chirality defined over the entire momentum space. 
At first glance, this Hamiltonian appears to violate particle number conservation, 
suggesting a breaking of the vector $\mathrm{U}(1)_V$ symmetry. 
However, we have confirmed the existence of charges that flow to the vector $\mathrm{U}(1)_V$ and axial $\mathrm{U}(1)_A$ symmetries in the continuum limit, 
and these charges commute with the Hamiltonian.
Naturally, this Hamiltonian does not violate the Nielsen--Ninomiya's theorem.  
This is because the eigenvalues of $Q_V$ are not generally integers in the momentum space, and $Q_V$ and $Q_A$ are non-commutative, preventing their simultaneous gauging.  

%%%%%%%%%%%%%%%%%%%%%%%%%%%%%%%%%%%%%%%%%%%%%%%%%%%%
%%%%%%%%%%%%%%%%%%%%%%%%%%%%%%%%%%%%%%%%%%%%%%%%%%%%
%%%%%%%%%%%%%%%%%%%%%%%%%%%%%%%%%%%%%%%%%%%%%%%%%%%%
%%%%%%%%%%%%%%%%%%%%%%%%%%%%%%%%%%%%%%%%%%%%%%%%%%%%
\section{Application to Symmetric Mass Generation}
\label{sec:SMG}
A promising method to realize chiral gauge theories on the lattice is the 
{Symmetric Mass Generation} (SMG) mechanism~\cite{Wang:2022ucy,Razamat:2020kyf,Tong:2021phe}. 
The SMG mechanism introduces suitable interaction terms that gap out a system 
while preserving its symmetries, without relying on any fermion bilinear terms. 
Although such interaction terms are typically irrelevant in the weak-coupling regime, 
it has been demonstrated numerically that they can become relevant in the strong-coupling regime~\cite{Zeng:2022grc}.

Given a fermionic system with a spacetime-internal symmetry group $G$, the following three conditions must be satisfied to realize SMG,
\begin{enumerate}[(i)]
  \item The theory must be free of any anomalies associated with the symmetry $G$.
  \item The symmetry $G$ must forbid all fermion bilinear mass terms.
  \item The condition of {instanton saturation} must be fulfilled.
\end{enumerate}
Here, condition~(iii) will become essential 
when (either dynamical or classical back-ground) gauge fields under symmetry $G$ 
with nontrivial topological (instanton) number are introduced,
which generate chiral zero modes. 
Consequently, in order to construct a well-defined theory, 
the chiral zero mode in the mirror sector must be saturated 
through the interaction term. We refer to this requirement as the \textit{instanton saturation} condition.

Based on the discussion in Sec.~\ref{sec:Eigenstates-Axial-charge}, it is possible to define Weyl fermions with quantized axial charges. 
In this section, we consider two examples—the $1^4(-1)^4$ model~\cite{Kikukawa:2017gvk} and the 3-4-5-0 model~\cite{Wang:2013yta,Wang:2018ugf,Wang:2022fzc,Zeng:2022grc,Onogi:2025tev}—to explore how the SMG mechanism can be realized while preserving the $\mathrm{U}(1)_A$ gauge symmetry generated by the axial charge operator $Q_A$.

%%%%%%%%%%%%%%%%%%%%%%%%%%%%%%%%%%%%%%%%%%%%%%%%%%%%
%%%%%%%%%%%%%%%%%%%%%%%%%%%%%%%%%%%%%%%%%%%%%%%%%%%%
\subsection{$(1)^4(-1)^4$ Model}
\label{subsec:1414-model}

In this subsection, we construct the $(1)^4(-1)^4$ model using  Weyl fermions possessing quantized axial charges. 
This model was previously studied in Ref.~\cite{Kikukawa:2017gvk} in the context of overlap fermion systems. It is expected that this model can acquire a gap without breaking any anomaly-free symmetries, as it can be regarded as an effective model for the chiral limit of the eight-flavor one-dimensional Majorana chain with reduced $\mathrm{SO}(6)$ symmetry (see Ref.~\cite{Kikukawa:2017gvk}).

This model is a four-flavor axial gauge theory with $\mathrm{Spin}(6)\ (\mathrm{SU}(4))$ flavor symmetry. 
The Hamiltonian of the model is given by
\begin{align}
    \label{eq:Def:4-flavor-Staggered-Hamiltonian}
    H = \i \sum_{f=1}^{4} \sum_{j=1}^{N} \biggl[ &
        \Psi_{f,L,j}^{\dagger} \frac{1}{2} (\nabla + \nabla^\dagger) \Psi_{f,L,j} 
        - \Psi_{f,R,j}^{H\dagger} \frac{1}{2} (\nabla + \nabla^\dagger) \Psi_{f,R,j} \notag \\
        & - \Psi_{f,R,j} \frac{1}{2} \nabla \nabla^\dagger \Psi_{f,L,j} 
        - \Psi_{f,L,j}^{H\dagger} \frac{1}{2} \nabla \nabla^\dagger \Psi_{f,R,j}^{\dagger} 
    \biggr] \ ,
\end{align}
which can be derived from the staggered fermion Hamiltonian. 
The Weyl fermions $\Psi_{f,L,j}$ and $\Psi_{f,R,j}$ are assumed to transform under the four-dimensional irreducible spinor representation of $\mathrm{SO}(6)$, (i.e.,$\underline{\mathrm{4}}$,and $\bar{4}$).
As discussed above, the Hamiltonian commutes 
with the following flavor-resolved vector and axial charge operators,
\begin{align}
    Q_{V_f} &= \frac{1}{2} \sum_{j=1}^{N} \biggl[
        \Psi_{f,L,j}^{\dagger} \Psi_{f,L,j}
        + \Psi_{f,R,j}^{\dagger} \Psi_{f,R,j}
        - \Psi_{f,L,j}^{\dagger} \Psi_{f,R,j}
        - \Psi_{f,R,j}^{\dagger} \Psi_{f,L,j}
        - 2 \notag \\
        &\qquad\quad - \frac{1}{2} \Bigl\{ \Psi_{f,L,j} - \Psi_{f,L,j}^{\dagger} + \Psi_{f,R,j} - \Psi_{f,R,j}^{\dagger} \Bigr\}
        \Bigl\{ \Psi_{f,L,j+1} + \Psi_{f,L,j+1}^{\dagger} + \Psi_{f,R,j+1} + \Psi_{f,R,j+1}^{\dagger} \Bigr\}
    \biggr] \ , \\
    Q_{A_f} &= \sum_{j=1}^{N} \left[
        \Psi_{f,L,j}^{\dagger} \Psi_{f,L,j}
        - \Psi_{f,R,j}^{\dagger} \Psi_{f,R,j}
    \right] \ ,
\end{align}
where $Q_{V_f}$ serves as the fermion number operator in the continuum limit. The subscript $f$ labels the flavor index.

Using these operators, we define the total axial and vector $\mathrm{U}(1)$ charges as follows,
\begin{align}
    Q_A &= Q_{A_1} + Q_{A_2} + Q_{A_3} + Q_{A_4} \ , \\
    Q_V &= Q_{V_1} + Q_{V_2} + Q_{V_3} + Q_{V_4} \ ,
\end{align}
see also Table~\ref{tab:charge-assignments-14-14-continuum}.
These charge assignments are summarized in Table~\ref{tab:charge-assignments-14-14-continuum}. Since $[Q_V, Q_A] \neq 0$, the vector charge $Q_V$ is anomalous, whereas the axial charge $Q_A$ and the $\mathrm{Spin}(6)\ (\mathrm{SU}(4))$ symmetry can be anomaly-free.
\begin{table}[t]
    \centering
    \begin{tabular}{|c||c|c|c|} \hline
        & $\mathrm{U}(1)_A$ & $\mathrm{U}(1)_V$ & $\mathrm{Spin}(6)\ (\mathrm{SU}(4))$   \\ \hline
        $(\Psi_{1,L}^{\dagger}, \Psi_{1,R}^{\dagger})$ & (1, -1) & (1, 1) & $(\underline{\mathrm{4}}, \bar{4})$  \\ \hline
        $(\Psi_{2,L}^{\dagger}, \Psi_{2,R}^{\dagger})$ & (1, -1) & (1, 1) & $(\underline{\mathrm{4}}, \bar{4})$  \\ \hline
        $(\Psi_{3,L}^{\dagger}, \Psi_{3,R}^{\dagger})$ & (1, -1) & (1, 1) & $(\underline{\mathrm{4}}, \bar{4})$  \\ \hline
        $(\Psi_{4,L}^{\dagger}, \Psi_{4,R}^{\dagger})$ & (1, -1) & (1, 1) & $(\underline{\mathrm{4}}, \bar{4})$  \\ \hline
    \end{tabular}
    \caption{Charge assignments for the $(1)^4(-1)^4$ model in the continuum.}
    \label{tab:charge-assignments-14-14-continuum}
\end{table}

Due to these symmetries, any fermion bilinear mass terms are forbidden. In order to gap out the system, one must introduce suitable interaction terms that explicitly break the anomalous $\mathrm{U}(1)_V$ symmetry while preserving the axial and flavor symmetries. An example of such interactions is given by
\begin{align}
    \Delta H = \sum_{j} \[ \Delta  _1 + \Delta  _2  \] \ ,
\end{align}
where
\begin{align}
    \Delta  _1 \propto \sum_{f=1}^{4}  \left(\sum_{f=1}^{4} \Psi_{f,L,j}^{\dagger} \Psi_{f,R,j}^{\dagger} \right)^2 \ , \ \Delta  _2 \propto \left(  \sum_{f=1}^{4} \Psi_{f,L,j}^{} \Psi_{f,R,j}^{} \right)^2 \ ,
\end{align}
which commutes with $Q_A$ but not with $Q_V$, and is invariant under the $\mathrm{Spin}(6)\ (\mathrm{SU}(4))$ flavor symmetry.
The square of $ \Delta_1$ and $ \Delta _2$ play a role of 't~Hooft vertexes, satisfying instanton saturation~(condition(iii)).

%%%%%%%%%%%%%%%%%%%%%%%%%%%%%%%%%%%%%%%%%%%%%%%%%%%%
%%%%%%%%%%%%%%%%%%%%%%%%%%%%%%%%%%%%%%%%%%%%%%%%%%%%
%%%%%%%%%%%%%%%%%%%%%%%%%%%%%%%%%%%%%%%%%%%%%%%%%%%%

\subsection{3-4-5-0 model}
\label{subsec:3450-model}

Following the previous section, 
we begin by considering the four-flavor massless Dirac Hamiltonian,
\begin{align}
    \label{eq:Def:3-4-5-0-Hamiltonian}
    H =  \i \sum_{f=1}^{4} \sum_{j=1}^{N} \biggl[ &
        \Psi_{f,L,j}^{H\dagger} \frac{1}{2}(\nabla + \nabla^\dagger) \Psi_{f,L,j}^{H}
        - \Psi_{f,R,j}^{H\dagger} \frac{1}{2}(\nabla + \nabla^\dagger) \Psi_{f,R,j}^{H} \notag \\
        & - \Psi_{f,R,j}^{H} \frac{1}{2} \nabla \nabla^\dagger \Psi_{f,L,j}^{H}
        - \Psi_{f,L,j}^{H\dagger} \frac{1}{2} \nabla \nabla^\dagger \Psi_{f,R,j}^{H\dagger}
    \biggr] \ .
\end{align}
We then investigate how to construct the 3-4-5-0 model~\cite{Wang:2013yta,Wang:2018ugf,Wang:2022fzc,Zeng:2022grc,Onogi:2025tev} 
by introducing suitable interaction terms that gap out one chirality of each flavor, 
specifically $\Psi_{1,L,j}, \Psi_{2,L,j}, \Psi_{3,R,j}, \Psi_{4,R,j}$, leaving the chiral fermions $\Psi_{1,R,j}, \Psi_{2,R,j}, \Psi_{3,L,j}, \Psi_{4,L,j}$ as the low-energy degrees of freedom.

The Hamiltonian preserves four $\mathrm{U}(1)$ symmetries, each associated with the following flavor-resolved charge operators,
\begin{align}
    Q_{V_f} &= \frac{1}{2} \sum_{j=1}^{N} \biggl[
        \Psi_{f,L,j}^{H\dagger} \Psi_{f,L,j}^{H}
        + \Psi_{f,R,j}^{H\dagger} \Psi_{f,R,j}^{H}
        - \Psi_{f,L,j}^{H\dagger} \Psi_{f,R,j}^{H}
        - \Psi_{f,R,j}^{H\dagger} \Psi_{f,L,j}^{H} - 2 \notag \\
        &\quad - \frac{1}{2} \left\{
            \Psi_{f,L,j}^{H} - \Psi_{f,L,j}^{H\dagger}
            + \Psi_{f,R,j}^{H} - \Psi_{f,R,j}^{H\dagger}
        \right\}
        \left\{
            \Psi_{f,L,j+1}^{H} + \Psi_{f,L,j+1}^{H\dagger}
            + \Psi_{f,R,j+1}^{H} + \Psi_{f,R,j+1}^{H\dagger}
        \right\}
    \biggr] \ , \\
    Q_{A_f} &= \sum_{j=1}^{N} \left[
        \Psi_{f,L,j}^{H\dagger} \Psi_{f,L,j}^{H}
        - \Psi_{f,R,j}^{H\dagger} \Psi_{f,R,j}^{H}
    \right] \ .
\end{align}
The $\( \mathrm{U}(1)\)^4 = \mathrm{U}(1)_{a1} \times \mathrm{U}(1)_{a2} \times \mathrm{U}(1)_{v1} \times \mathrm{U}(1)_{v2}$ symmetry of the 3-4-5-0 model in Table~\ref{tab:charge-assignments-3450-continuum} 
is defined in terms of these charge operators as\footnote{
        In principle, the charge assignments of the $\mathrm{U}(1)$ gauge symmetry and the extra $\mathrm{U}(1)$ symmetry in this model can be uniquely determined, 
        both in the lattice and continuum theories, by the condition that anomaly matching is equivalent to the boundary fully gapping condition~\cite{Wang:2013yta,Wang:2018ugf}. 
In our lattice construction, however, the anomaly matching conditions for the $\mathrm{U}(1)$ gauge symmetry and the extra $\mathrm{U}(1)$ symmetry become trivial by construction, i.e., $[Q_{a1}, Q_{a2}] = 0$. As a result, the charge assignments are no longer uniquely fixed by anomaly considerations alone. 
In this work, we therefore choose the charge assignments so that they consistently reproduce those of the 3-4-5-0 model in the continuum limit. 
A discussion of anomaly cancellation in this model can also be found in Ref.~\cite{Li:2024dpq}.
},
\begin{align}
    Q_{a1} &= 3 Q_{A_1} + 4 Q_{A_2} - 5 Q_{A_3} - 0 Q_{A_4} \ , \\
    Q_{a2} &= 0 Q_{A_1} + 5 Q_{A_2} - 4 Q_{A_3} - 3 Q_{A_4} \ , \\
    Q_{v1} &= 3 Q_{V_1} + 4 Q_{V_2} - 5 Q_{V_3} - 0 Q_{V_4} \ , \\
    Q_{v2} &= 0 Q_{V_1} + 5 Q_{V_2} - 4 Q_{V_3} - 3 Q_{V_4} \ .
\end{align}

Although the theory remains vector-like at this stage, any fermion bilinear mass terms are strictly forbidden by the $\mathrm{U}(1)_{a1} \times \mathrm{U}(1)_{a2}$ symmetry. 
This model exhibits 't~Hooft anomalies between the axial and vector symmetries,
\begin{align}
    [Q_{a,s}, Q_{v,s'}] \neq 0 \ ,
\end{align}
where $s,s' = 1,2$.
It should be noted that the 't~Hooft anomaly cancellation for $\mathrm{U}(1)_{a1} \times \mathrm{U}(1)_{a2}$ is trivial in our setup by construction.
Therefore, to gap out a specific set of chiral fermions while preserving this symmetry, 
one must explicitly break the anomalous $\mathrm{U}(1)_{v1} \times \mathrm{U}(1)_{v2}$ symmetry through appropriate multi-fermion interactions. 
Such interactions are given by,
\begin{align}
    \label{eq:3450-interaction1}
    \Delta_{H1} &= \sum_{j} \[ \Delta_{1} +  \text{h.c.} \]  \ , \\
    \label{eq:3450-interaction2}
    \Delta_{H2} &=  \sum_{j} \[ \Delta_{2}  + \text{h.c.} \] \ ,
\end{align}
where
\begin{align}
    \Delta_{1} & \propto {\Psi}_{1,L,j}
    \left({\Psi}_{2,L,j}^\dagger {\Psi}_{2,L,j+1}^\dagger\right)
    {\Psi}_{3,R,j}
    \left({\Psi}_{4,R,j} {\Psi}_{4,R,j+1} \right) \ , \\
    \Delta_{2} & \propto \left({\Psi}_{1,L,j} {\Psi}_{1,L,j+1}\right)
    {\Psi}_{2,L,j}
    \left({\Psi}_{3,R,j}^\dagger {\Psi}_{3,R,j+1}^\dagger\right)
    {\Psi}_{4,R,j} \ .
\end{align}
These interaction terms do not commute with the vector charge operators $Q_{v,f}$, but they do commute with the axial charges $Q_{a,f}$. 
Importantly, this commutation property holds not only at $k = 0$ but throughout the entire momentum space. 
This feature sharply contrasts with the construction in Ref.~\cite{Xu:2025hfs}, 
where the 3-4-5-0 model is realized starting from a two-flavor Dirac fermion system. 
In that setup, one of the two required charge operators does not commute with the interaction terms corresponding to Eqs. \eqref{eq:3450-interaction1},\eqref{eq:3450-interaction2} in the full momentum space.
Note that the combination of the interactions $\Delta_1$ and $\Delta_2$ such as $\Delta_1 ^\dagger \( \Delta_2\)^2 $
generates an appropriate 't~Hooft vertex,
and therefore satisfies the instanton saturation condition (condition (iii)).

Whether this setup indeed realizes the 3-4-5-0 model by successfully gapping out the intended chiral fermions remains a subject for future numerical investigation.
\begin{table}[tb]
    \centering
    \begin{tabular}{|c||c|c|c|c|} \hline
        & $\mathrm{U}\(1\)_{a1}$ & $\mathrm{U}\(1\)_{a2}$ & $\mathrm{U}\(1\)_{v1}$ & $\mathrm{U}\(1\)_{v2}$  \\ \hline
        $\(\Psi_{1,L}^{\dagger}, \Psi_{1,R}^{\dagger}\)$ & (3,-3) & (0,0) & (3,3) & (0,0) \\ \hline
        $\(\Psi_{2,L}^{\dagger}, \Psi_{2,R}^{\dagger}\)$ & (4,-4) & (5,-5) & (4,4) & (5,5)\\ \hline
        $\(\Psi_{3,L}^{\dagger}, \Psi_{3,R}^{\dagger}\)$ & (-5,5) & (-4,4) & (-5,-5) & (-4,-4) \\ \hline
        $\(\Psi_{4,L}^{\dagger}, \Psi_{4,R}^{\dagger}\)$ & (0,0) & (-3,3) & (0,0) & (-3,-3)\\ \hline
    \end{tabular}
    \caption{Charge assignments for the 3-4-5-0 model in the continuum.}
    \label{tab:charge-assignments-3450-continuum}
\end{table}

%%%%%%%%%%%%%%%%%%%%%%%%%%%%%%%%%%%%%%%%%%%%%%%%%%%%
%%%%%%%%%%%%%%%%%%%%%%%%%%%%%%%%%%%%%%%%%%%%%%%%%%%%
\section{Conclusion}
\label{sec:Conclusion}
In this work, we have reconstructed two lattice charge operators,
$Q_V$ and $Q_A$, that commute with the Hamiltonian of the 1+1~D staggered fermion,
using the Wilson fermion formalism.
We have shown that these charges flow to the generators of the vector $\mathrm{U}(1)_V$ 
and axial $\mathrm{U}(1)_A$ symmetries, respectively, in the continuum limit.
Furthermore, we have constructed the Hamiltonian using the eigenstates of $Q_A$,
which ensures that the axial $\mathrm{U}(1)_A$ symmetry is preserved.
Although the resulting Hamiltonian may appear to explicitly break
the naive vector $\mathrm{U}(1)_V$ symmetry, it is, by construction,
commutable with $Q_V$ and thus retains the $\mathrm{U}(1)_V$ symmetry
in the continuum limit.
While this result might seem trivial,
it is nevertheless significant that the Hamiltonian remains compatible
with $Q_V$ even when explicitly constructed to respect axial symmetry on the lattice.
One key advantage of respecting $Q_A$ is that it acts locally on operators,
which implies that it can be gauged. This suggests the possibility
of formulating chiral $\mathrm{U}(1)_A$ gauge theories on the lattice at least
in certain restricted settings.

As an application,
we have discussed how to formulate the Symmetric Mass Generation (SMG) mechanism
in the context of free chiral $\mathrm{U}(1)_A$ gauge theories (without dynamical gauge fields),
using the $1^4(-1)^4$ model and the 3-4-5-0 model as examples.
Since the Hamiltonian is constructed from fermions with well-defined
and constant integer-valued chirality over the entire momentum space,
the interactions introduced to achieve gapping can be consistently defined
without violating the symmetries that must be preserved.
However, in the case of the 3-4-5-0 model,
further numerical investigation is required
to confirm whether the SMG mechanism genuinely occurs within this setup.
The Hamiltonian we used contains terms corresponding to the Wilson mass,
which generically mix chiralities. It is nontrivial
whether one can successfully gap out one chiral fermion while
keeping the other massless solely through interaction terms in this framework.
We leave this question as a direction for future work.

    Finally, one of the motivations for adopting the Hamiltonian approach is 
    its compatibility with quantum simulation platforms.
    In particular, our formulation may serve as a theoretical foundation for future implementations using ultra cold atoms,
    which could offer a promising route to overcoming the sign problem in strongly correlated systems~\cite{Zohar:2015hwa}.

%%%%%%%%%%%%%%%%%%%%%%%%%%%%%%%%%%%%%%%%%%%%%%%%%
%%%%%%%%%%%%%%%%%%%%%%%%%%%%%%%%%%%%%%%%%%%%%%%%%
\acknowledgments
We sincerely thank Tetsuya Onogi for the many fruitful discussions, which were instrumental in shaping the direction and progress of this work.
The work is supported in part by JST SPRING, Grant Number JP- MJSP2138.

%%%%%%%%%%%%%%%%%%%%%%%%%%%%%%%%%%%%%%%%%%%%%%%%%
%%%%%%%%%%%%%%%%%%%%%%%%%%%%%%%%%%%%%%%%%%%%%%%%%%
\appendix

%%%%%%%%%%%%%%%%%%%%%%%%%%%%%%%%%%%%%%%%%%%%%%%%%%%%
%%%%%%%%%%%%%%%%%%%%%%%%%%%%%%%%%%%%%%%%%%%%%%%%%%%%

\section{Onsager algebra}
\label{sec:Onsager} 

We briefly introduce the Onsager algebra~\cite{Onsager:1943jn}; for details, see Section D of Ref.~\cite{Chatterjee:2024gje}.

Let us define the operators
\begin{align}
    \label{eq:Onsager-Gn}
    G_n &= \frac{\i}{2} \sum_j \left( a_j a_{j+n} - b_j b_{j+n} \right) \ , \\
    \label{eq:Onsager-Qn}
    Q_n &= \frac{\i}{2} \sum_j a_j b_{j+n} \ , \quad n \in \mathbb{Z} \ .
\end{align}
These operators commute with the Hamiltonian~\eqref{eq:Def:Staggered-Hamiltonian} and satisfy the following closed algebra,
\begin{align}
    [Q_n, Q_m] &= \i G_{m-n} \ , \\
    [G_n, G_m] &= 0 \ , \\
    [Q_n, G_m] &= 2\i \left( Q_{n-m} - Q_{n+m} \right) \ .
\end{align}

The charge operators $Q_V$ and $Q_A$, which commute with the staggered fermion Hamiltonian~\eqref{eq:Def:Staggered-Hamiltonian}, can be expressed using the Onsager algebra as~\cite{Chatterjee:2024gje}
\begin{equation}
    Q_V = Q_0 \ , \quad Q_A = Q_1 \ , \quad \tilde{Q}_A = \frac{1}{2} \left( Q_1 + Q_{-1} \right) \ .
\end{equation}
While $Q_n = T_b^n Q_0 T_b^{-n}$ possess integer eigenvalues, the $G_n$ operators generally do not.
Here, the third charge operator $\tilde{Q}_A$ is an axial operator defined to satisfy $[H,\tilde{Q}_A ] = [Q_V, \tilde{Q}_A] = 0$. Explicitly, it is given by
\begin{align}
    \label{eq:Def:unquantized-axial-operator}
    \tilde{Q}_A &= \frac{1}{2} \left( Q_A + e^{\frac{\i \pi}{2} Q_V} Q_A e^{-\frac{\i \pi}{2} Q_V} \right) \notag \\
    &= \frac{1}{2} \sum_{j=1}^{2N} \left( c_j^\dagger c_{j+1} - c_j c_{j+1}^\dagger \right) \notag \\
    &\equiv \sum_{j=1}^{2N} \tilde{q}^A_{j+\frac{1}{2}} \ .
\end{align}
As a consequence, the eigenvalues of $\tilde{Q}_A$ are not quantized.

%%%%%%%%%%%%%%%%%%%%%%%%%%%%%%%%%%%%%%%%%%%%%%%%%%
%%%%%%%%%%%%%%%%%%%%%%%%%%%%%%%%%%%%%%%%%%%%%%%%%%
\section{The chiral charge in the Wilson Hamiltonian}
\label{OV-Wilson-Chiralcharge}

In 1+1 dimensions, the Wilson--Dirac operator defined by Eq.~\eqref{Wilson-E} satisfies
\begin{equation}
    \mathcal{D}_W^\dagger (m=1)\mathcal{D}_W(m=1) = 1 \ ,
\end{equation}
which implies that the Wilson--Dirac operator at $m=0$ can be expressed as
\begin{align}
    \mathcal{D}_W (m=0) &= 1 +  \frac{\mathcal{D}_W(m=1)}{\sqrt{\mathcal{D}_W^\dagger (m=1)\mathcal{D}_W(m=1)}} 
    =: \mathcal{D}_\OV (m=1) \ .
\end{align}
This transformation introduces the overlap Dirac operator $\mathcal{D}_\OV$~\cite{Creutz:2001wp,Hayata:2023zuk,Hayata:2023skf}, 
which satisfies relations analogous to the Ginsparg--Wilson relation~\cite{Ginsparg:1981bj},
\begin{align}
    \gamma^5 \mathcal{D}_\OV + \mathcal{D}_\OV \gamma^5 &= \mathcal{D}_\OV \gamma^5 \mathcal{D}_\OV \ , \notag \\  
    \gamma^0 \mathcal{D}_\OV + \mathcal{D}_\OV \gamma^0 &= \mathcal{D}_\OV \gamma^0 \mathcal{D}_\OV \ , \notag\\  
    \gamma^5 \mathcal{D}_\OV &= \mathcal{D}_\OV^\dagger \gamma^5 \ , \notag\\  
    \gamma^0 \mathcal{D}_\OV &= \mathcal{D}_\OV^\dagger \gamma^0 \ .
    \label{eq:GW-relations}
\end{align}
Therefore, in 1+1 dimensions, since the Wilson--Dirac operator obeys a Hamiltonian version of the Ginsparg--Wilson relation~\cite{Ginsparg:1981bj}, 
the Hamiltonian of the Wilson fermion can be regarded as equivalent to that of the overlap fermion~\cite{Creutz:2001wp,Hayata:2023zuk,Hayata:2023skf}.

Using the overlap Dirac operator $\mathcal{D}_\OV$, one can define the chiral charge operator~\cite{Hayata:2023zuk} as
\begin{align}
    \label{eq:Def:chiral-charge-operator}
    Q_\chi = \sum_j \psi_j^\dagger \gamma^5 \left(1 - \frac{1}{2} \mathcal{D}_\OV \right) \psi_j \ ,
\end{align}
which commutes with the overlap Hamiltonian $H_\OV$,
\begin{align}
    [H_W , Q_\chi] &= 0 \ .
\end{align}
The chiral charge operator $Q_\chi$ can also be rewritten in terms of the one-component staggered fermion $c_j$ as
\begin{align}
    Q_\chi &= \sum_j \psi_j^\dagger \gamma^5 \left(1 - \frac{1}{2} \mathcal{D}_W \right) \psi_j 
    = \frac{1}{2} \sum_j \left( c_j^\dagger c_{j+1} - c_j c_{j+1}^\dagger \right) 
    = \tilde{Q}_A \ .
\end{align}
Namely, although this charge satisfies $[H_\W, Q_\chi] = [Q_V, Q_\chi] = 0$, its eigenvalues are unquantized~\cite{Chatterjee:2024gje}.

The momentum-space representation makes it evident that the eigenvalues of the chiral charge operator become unquantized. 
Let $\ket{\psi_k^{\chi}}$ denote a simultaneous eigenstate of the Wilson Hamiltonian $H_\W$ and the chiral charge operator ${Q}_\chi$, with respective eigenvalues $\epsilon_k$ and $q_k$, satisfying
\begin{align}
    H_{\text{W}} \ket{\psi_k^{\chi}} &= \epsilon_k \ket{\psi_k^{\chi}}\ , \\
    {Q}_\chi \ket{\psi_k^{\chi}} &= q_k \ket{\psi_k^{\chi}} \ .
\end{align}
Then, it can be straightforwardly shown from Eqs.~\eqref{eq:GW-relations} that the following identity holds,
\begin{align}
    \label{eq:relation-EnergyandChirality}
    \frac{1}{4} \epsilon_k^2 + q_k^2 = 1 \ .
\end{align}
This relation indicates that the chirality defined by $Q_\chi$ 
depends on the momentum and generally takes non-integer values. 
As a result, chiral projections based on the Ginsparg--Wilson relation, 
which are well-defined in the path-integral formalism 
using the corresponding Dirac operator, cannot be naively implemented 
in the Hamiltonian formalism.

%%%%%%%%%%%%%%%%%%%%%%%%%%%%%%%%%%%%%%%%%%%%%%%%%%%%
%%%%%%%%%%%%%%%%%%%%%%%%%%%%%%%%%%%%%%%%%%%%%%%%%%%%

%%%%%%%%%%%%%%%%%%%%%%%%%%%%%%%%%%%%%%%%%%%%%%%%%%%%
%%%%%%%%%%%%%%%%%%%%%%%%%%%%%%%%%%%%%%%%%%%%%%%%%%%%

%\newpage
\bibliographystyle{utphys}
\bibliography{refer}

\end{document}